\documentclass[twocolumn,aps,showpacs,floatfix,prc]{revtex4}
\usepackage[dvips]{epsfig}
\usepackage{subeqn}

\begin{document}

\title{Primordial lithium abundance problem of BBN and baryonic density in the universe}

\author{ Vinay Singh$^{1,\S,\dagger}$, Joydev Lahiri$^{2,\S}$, Debasis Bhowmick$^{3,\S}$ and D. N. Basu$^{4,\S,\dagger}$ }

\affiliation{$^{\S}$Variable Energy Cyclotron Centre, 1/AF Bidhan Nagar, Kolkata 700 064, India }
\affiliation{$^\dagger$Homi Bhabha National Institute, Training School Complex, Anushakti Nagar, Mumbai 400 085, India }

\email[E-mail 1: ]{vsingh@vecc.gov.in}
\email[E-mail 2: ]{joy@vecc.gov.in}
\email[E-mail 3: ]{dbhowmick@vecc.gov.in}
\email[E-mail 4: ]{dnb@vecc.gov.in}

\date{\today }

\begin{abstract}

    Prediction of the primordial abundances of elements in the big-bang nucleosynthesis (BBN) is one of the three strong evidences for the big bang model. Precise knowledge of the baryon-to-photon ratio of the Universe from observations of the anisotropies of cosmic microwave background radiation has made the Standard BBN a parameter-free theory. Although, there is a good agreement over a range of nine orders of magnitude between abundances of light elements deduced from observations and calculated in primordial nucleosynthesis, there remains a yet-unexplained discrepancy of $^7$Li abundance higher by a factor of $\sim 3$ when calculated theoretically. The primordial abundances depend on the astrophysical nuclear reaction rates and on three additional parameters, the number of light neutrino flavours, the neutron lifetime and the baryon-to-photon ratio in the universe. The effect of the modification of thirty-five reaction rates on light element abundance yields in BBN was investigated earlier by us. In the present work we have incorporated the most recent values of neutron lifetime and the baryon-to-photon ratio and further modified $^3$He($^4$He,$\gamma$)$^7$Be reaction rate which is used directly for estimating the formation of $^7$Li as a result of $\beta^+$ decay as well as the reaction rates for t($^4$He,$\gamma$)$^7$Li and d($^4$He,$\gamma$)$^6$Li. We find that these modifications reduce the theoretically calculated abundance of $^7$Li by $\sim 12\%$.

\vspace{0.2cm}
\end{abstract}

\maketitle

\noindent
\section{Introduction}
\label{section1}

    The Hubble expansion of the Universe, the Cosmic Microwave Background Radiation (CMBR) and the big bang nucleosynthesis (BBN) are the three signatures of the big bang model. These are supported by a large number of observational evidences. The BBN, which predicts the primordial abundances of the light elements such as D, $^{3,4}$He and $^{6,7}$Li, took place just a few moments after the big-bang \cite{Ho64} and the universe evolved very rapidly allowing only the synthesis of the lightest nuclides. In addition to these stable nuclei some unstable, or radioactive, isotopes like tritium or ${^3}$H and $^{7,8}$Be were also produced during the BBN. These unstable isotopes either decayed or fused with other nuclei to make one of the stable isotopes. It lasted for only about seventeen minutes (during the period from three to about twenty minutes from the beginning of space expansion) and after that, the temperature and density of the universe fell below that which is required for nuclear fusion and prevented elements heavier than beryllium to form while at the same time allowing unburned light elements, such as deuterium, to exist. 
    
    Although, there is a good agreement between primordial abundances of D and $^{3,4}$He deduced from observations and from primordial nucleosynthesis calculations, but that of $^{6,7}$Li are off by quite large factors. The predictions of the standard BBN theory depend on the astrophysical nuclear reaction rates and on three additional parameters, {\it viz.}, the number of flavours of light neutrinos, the neutron lifetime and the baryon-to-photon ratio in the Universe. The observations by the Wilkinson Microwave Anisotropy Probe [WMAP] \cite{WMAP,WMAP1} and the Planck \cite{Planck,Planck1} space missions enabled precise extraction of the baryon-to-photon ratio of the Universe. The weak reaction rates involved in n-p equilibrium come from the standard theory of the weak interaction. These are calculated \cite{Di82} with neutron lifetime as the only experimental input, whose recent experimental value, 880.3$\pm$1.1 s \cite{Ol14}, may be further updated \cite{Wi11,Yo14}, that influence the production of $^{4}$He \cite{Ma05}. In the past, sensitivity to several parameters and physics inputs in the BBN model were investigated \cite{No00,Cy02,Cy04,Cy08,Se04,Io09,Fu10}.

    The most important inputs for modeling the BBN and stellar evolution are the nuclear reaction rates $<\sigma v>$ in the reaction network calculations, where $\sigma$ is the nuclear-fusion cross section and $v$ is the relative velocity between the participant nuclides. These low energy fusion cross sections can only be obtained from laboratory experiments, some of which are not as well known \cite{No00,Cy02,Cy04,Cy08,Se04}. However, $v$ is well described by a Maxwellian velocity distribution for a given temperature $T$. Several factors influence the measured values of the cross sections and the theoretical estimates of the thermonuclear reaction rates depend on the various approximations used. In the network calculations one needs to account for the Maxwellian-averaged thermonuclear reaction rates and the difference \cite{Fo88,An99} in these rates affects the description of elemental synthesis in the BBN or in stellar evolution. 
      
    In this work, we consider the effects of the nuclear reaction rates, neutron lifetime and the baryon-to-photon ratio on the primordial abundances of elements. The effect of the modification of thirty-five reaction rates on light element abundance yields in BBN was investigated earlier by us. We have used the most recent values of neutron lifetime and baryon-to-photon ratio and further modified reaction rates for d($^4$He,$\gamma$)$^6$Li, t($^4$He,$\gamma$)$^7$Li and $^3$He($^4$He,$\gamma$)$^7$Be, which is used directly for estimating the formation of $^7$Li as a result of $\beta^+$ decay, by the most recent rate equations in the temperature ranges up to 5$T_9$ (in units of $10^9$ K) \cite{Du17}. We have studied light element abundance yields as functions of evolution time or temperature.

\noindent
\section{Brief thermal history of the early universe}
\label{section2}

    The nucleosynthesis calculation requires the knowledge of time evolutions of the baryonic density and temperature. These may be obtained from the rate of expansion of the universe and thermodynamic considerations. The geometry of the universe is described by the Friedmann-Robertson-Walker (FRW) metric given by

\begin{equation}
 ds^2 = dt^2 - a^2(t) \Big( \frac{dr^2}{1-k r^2} + r^2(d\theta^2 + sin\theta d\phi^2)  \Big)
\label{seqn1}
\end{equation}
\noindent
which assumes homogeneity and isotropy, where $a(t)$ is the scale factor, describing the expansion, and k = 0, $\pm$1 denotes the flat, closed or open universe, respectively. From Einstein equations one obtains

\begin{equation}
 H^2(t) = \Big(\frac{\dot a}{a}\Big)^2 = \frac{8\pi G (\rho_R + \rho_M)}{3} - \frac{k}{a^2} + \frac{\Lambda}{3}
\label{seqn2}
\end{equation}
\noindent
where $H(t)$ is the Hubble parameter, $G$ is the gravitational constant, $\rho_M$ and $\rho_R$ are the matter and radiation densities respectively and $\Lambda$ is the cosmological constant. When considering the density components of the universe, it is convenient to consider the critical density $\rho_C=\frac{3H_0^2}{8\pi G}$ for a flat (Euclidean) space corresponding to $k = 0$, $\Lambda=0$ in Eq.(2).

    As is well known, during the early stages of expansion the matter density (dark and baryonic) $\rho_M \propto a^{-3}$, while the radiation density $\rho_R \propto a^{-4}$. During the BBN epoch, when $a$ is $\approx 10^{-8}$ times the present value, $H(t)$ is governed solely by relativistic particles while the matter density, cosmological constant and curvature terms play no role. In this case Eq.(2) takes the form

\begin{equation}
 \Big( \frac{\dot a}{a}\Big)^2 = \frac{8\pi G }{3}a_R \frac{g_*(T)}{2} \times T^4    
\label{seqn3}
\end{equation}
\noindent
where the Stefan-Boltzmann law $a_R T^4$ for the radiation energy density is used and $g_*$ is the effective spin factor which decreases whenever the temperature drops below a mass threshold for the particle-antiparticle annihilation of each species. During BBN, only e$^+$ and e$^-$ annihilates implying that the contributions to $g_*(T)$ come from photons, neutrinos and electrons/positrons before they annihilate. The released energy is shared among other particles which were in equilibrium with photons and baryons but not neutrinos as it occurs after their decoupling. Apart from thermonuclear reaction rates, the important quantities needed for BBN calculations are the photon/ion temperature, the neutrino temperature and the baryonic density as a function of time which is obtained by numerically solving Eq.(3) \cite{Wa67,Wa69} with the constraint that the entropy densities of neutrinos and photons+electrons stay separately constant during the adiabatic expansion \cite{Ka92,We08}. The baryonic density does not have at this epoch any effect on the rate of expansion of the universe but influences nucleosynthesis since higher density of nuclei induces a larger number of reactions taking place per unit time.

    The predictions of the standard BBN theory depend on the astrophysical nuclear reaction rates and on three additional parameters, the number of light neutrino flavours ($N_\nu$), the neutron lifetime ($\tau_n$) and the baryon-to-photon ratio ($\eta=n_B/n_\gamma$) in the universe \cite{Co95,Sh95}. In its standard $N_\nu=3.0$ form, BBN is a parameter-free theory because of the precise knowledge of the baryon-to-photon ratio of the Universe from the observations of the anisotropies of the CMBR. The big-bang cosmology relies on the Hubble expansion, CMBR and the BBN. The BBN examines back to earlier times of the universe and deals with nuclear and particle physics along with cosmology. Although Hubble expansion can also be used by some other alternative cosmological theories, the evidences of CMBR and BBN observations implies a universe which was very hot and dense at the very beginning. As described above, the FRW cosmological model is the standard scenario for the BBN theory. The facts that the solution to Einstein equations leads to a homogeneous and isotropic universe implying uniformity of the CMBR temperature, which is $T= 2.7277 \pm 0.002$ K across the sky, and the success of the standard BBN theory validate this approximation. It is possible to characterize the BBN in general, that the paradigm most frequently uses the Friedmann equation to relate the big-bang expansion rate, H, to the thermal properties of the particles present at that epoch. During expansion, the rates of the weak interactions that transform neutrons and protons, and the rates of the nuclear reactions that build up the complex nuclei are involved. The syntheses of light elements in the early universe is determined by the time during its expansion.  

\noindent
\section{Thermonuclear reaction rates and BBN reaction network}
\label{section3}

    The reactions which took place during BBN can be organized into two groups, {\it viz} the reactions that convert neutrons to protons and {\it vice versa}: n $+$ e$^+ \leftrightarrow$ p $+~\bar\nu_e$, p $+$ e$^- \leftrightarrow$ n $+~\nu_e$ and n $\leftrightarrow$ p $+$ e$^-+\bar\nu_e$ and the rest of the reactions. The first group can be expressed in terms of the mean neutron lifetime while the second group is determined by many different nuclear cross section measurements. The formation of deuterium begins in the process of p $+$ n $\leftrightarrow$ D $+~\gamma$. This reaction is exothermic with an energy difference of 2.2246 MeV, but since photons are 10$^9$ times more numerous than protons, the reaction does not proceed until the temperature of the expanding universe falls to about 0.3 MeV, when the photo-destruction rate is lower than the production rate of deuterons. When the deuteron formation starts some further reactions proceed to make ${^4}$He nuclei: D$~+ n \rightarrow {^3}$H$~+ \gamma$, ${^3}$H$~+~$p $\rightarrow {^4}$He $+$ $\gamma$, D $+$ p $\rightarrow {^3}$He $+~ \gamma$, ${^3}$He $+$ n $\rightarrow {^4}$He $+~\gamma$. Both light helium ${^3}$He and normal helium ${^4}$He are formed along with the ${^3}$H. Since helium nucleus binding energy is 28.3 MeV and more bound than the deuterons and the temperature has already fallen to 0.1 MeV, these reactions can be photo-reactions and only go one way. The four reactions: D $+$ D $\rightarrow {^3}$He $+$ n, D $+$ D $\rightarrow {^3}$H + p, ${^3}$He $+$ D $\rightarrow {^4}$He $+$ p, ${^3}$H $+$ D $\rightarrow {^4}$He $+$ n also produce ${^3}$He and ${^4}$He and they usually go faster since they do not involve the relatively slow process of photon emission. Eventually the temperature gets so low that the electrostatic repulsion of the deuterons and other charged particles causes the reactions to stop. The deuteron to proton ratio when the reactions stop is quite small, and essentially inversely proportional to the total density of protons and neutrons (to be precise, goes like the -1.6 power of the density). Almost all the neutrons in the universe end up in normal helium nuclei. For a neutron:proton ratio of 1:7 at the time of deuteron formation, 25$\%$ of the mass ends up in helium. Deuterium peaks around 100 seconds after the big-bang, and is then rapidly swept up into helium nuclei. A very few helium nuclei combine into heavier nuclei giving a small abundance of ${^7}$Li coming from the big-bang. ${^3}$H decays into ${^3}$He with a twelve year half-life so no ${^3}$H survives to the present, and ${^7}$Be decays into ${^7}$Li with about fiftythree day half-life and also does not survive. 
    
    Instead of cross sections $\sigma$, the nuclear reaction inputs to BBN take the form of thermal rates. The thermonuclear reaction rates are computed by averaging nuclear reaction cross sections over a Maxwell-Boltzmann distribution of energies. The Maxwellian-averaged thermonuclear reaction rate $<\sigma v>$ at some temperature $T$, is given by the following integral \cite{Bo08}:

\begin{equation}
 <\sigma v> = \Big[\frac{8}{\pi\mu (k_B T)^3 } \Big]^{1/2} \int \sigma(E) E \exp(-E/k_B T) dE,
\label{seqn4}
\end{equation}
\noindent
where $E$ is the centre-of-mass energy, $v$ is the relative velocity and $\mu$ is the reduced mass of the reactants. At low energies (far below Coulomb barrier) where the classical turning point is much larger than the nuclear radius, barrier penetrability can be approximated by $\exp(-2\pi\zeta)$ so that the charge induced cross section can be decomposed into

\begin{equation}
 \sigma(E) = \frac{S(E)\exp(-2\pi\zeta)}{E}
\label{seqn5}
\end{equation}
\noindent
where $S(E)$ is the astrophysical $S$-factor and $\zeta$ is the Sommerfeld parameter, defined by $\zeta = \frac{Z_1Z_2e^2}{\hbar v}$ where $Z_1$ and $Z_2$ are the charges of the reacting nuclei in units of elementary charge $e$. Except for narrow resonances, the $S$-factor $S(E)$ is a smooth function of energy, which is convenient for extrapolating measured cross sections down to astrophysical energies. In the case of a narrow resonance, the resonant cross section $\sigma_r(E)$ is generally approximated by a Breit-Wigner expression whereas the neutron induced reaction cross sections at low energies can be given by $\sigma(E)=\frac{R(E)}{v}$ \cite{Bl55} facilitating extrapolation of the measured cross sections down to astrophysical energies, where $R(E)$ is a slowly varying function of energy \cite{Mu10} and is similar to $S$-factor.

\noindent
\section{Observed primordial abundances}
\label{section4}

    After BBN, ${^4}$He is also produced by stars. Its primordial abundance is deduced from observations in ionized hydrogen regions of compact blue galaxies. Galaxies are thought to be formed by the agglomeration of such dwarf galaxies, in a hierarchical structure formation paradigm, which are hence considered as more primitive. To account for stellar production, ${^4}$He abundance deduced from observations is extrapolated to zero, followed by atomic physics corrections. Aver et al. \cite{Aver15} obtained its mass fraction to be $0.2449 \pm 0.0040$. 
    
    Deuterium can be destroyed after BBN throughout stellar evolution. Its primordial abundance is estimated from the observation of a few cosmological clouds on the line of sight of distant quasars at high redshift. Recently, Cooke et al. \cite{Cooke14} have reanalyzed existing data as well as made new observations that lead to D/H relative abundance of $(2.53 \pm 0.04) \times 10^{-5}$ with smaller uncertainties than previous estimates. 
    
    Contrary to ${^4}$He, ${^3}$He is both produced and destroyed in stars so that the evolution of its abundance as a function of time is not well known. Because of the difficulties of helium observations and the small ${^3}$He/${^4}$He ratio, ${^3}$He has only been observed in our Galaxy and its relative abundance is estimated to be $(1.1 \pm 0.2) \times 10^{-5}$ \cite{Ba02}.

    The BBN continued from about three to twenty minutes from the beginning of space expansion. Consequently, the temperature and density of the universe fell below the value which is required for nuclear fusion and thus prevented elements heavier than beryllium to form while at the same time allowed unburned light elements, such as deuterium, to exist. The heavier element nucleosynthesis takes place mainly in massive stars. During the evolution of galaxies, these stars explode as supernovae and release matter enriched in heavy elements into the interstellar medium. Accordingly, the abundances of heavier elements increase with time in stars. Therefore, the observed abundance of metals (elements heavier than helium) is an indication of age. To be explicit, the older ones have the lower metallicity. Hence, the primordial abundances are extracted from observations of objects with very small metallicity. After BBN $^7$Li can be both produced (spallation, AGB stars, novae) and destroyed (in the interior of stars). Very old stars can still be observed in the halo of our Galaxy since the life expectancy of stars with masses lower than the Sun is larger than the age of the universe. Lithium can be observed at the surface of these stars and its abundance was found to be remarkably constant independent of metallicity, below $\approx$0.1 of the solar metallicity. This constant plateau \cite{Sp82} of Li abundance was interpreted as corresponding to the BBN $^7$Li production. The thinness of plateau is an indication that surface Li depletion may not have been very effective and it should reflect the primordial value. The analysis of Sbordone et al. \cite{Sb10} provides $^7$Li/H $=(1.58^{+0.35}_{-0.28}) \times 10^{-10}$. 

\noindent
\section{ Calculations of abundances in primordial nucleosynthesis }
\label{section5}
     
    In our previous work \cite{Mi12}, we modified thirty-five Maxwellian-averaged thermonuclear reaction rates from Caughlan et al. \cite{Fo88} and Smith et al. \cite{Sm93} used in the Kawano/Wagoner BBN code \cite{Ka92} by the compilations of Angulo et al. \cite{An99} and Descouvemont et al. \cite{An04} which were meant to supersede the earlier compilations and studied its effect on the primordial abundances of elements. These reactions were d(p,$\gamma$)${^3}$He, d(d,n)${^3}$He, d(d,p)t, d($\alpha,\gamma$)${^6}$Li, t(d,n)${^4}$He, t($\alpha,\gamma$)${^7}$Li, ${^3}$He(n,p)t, ${^3}$He(d,p)${^4}$He, ${^3}$He(${^3}$He,2p)${^4}$He, ${^3}$He($\alpha,\gamma$)${^7}$Be, ${^4}$He($\alpha$n,$\gamma$)${^9}$Be, ${^4}$He($\alpha \alpha,\gamma$)${^{12}}$C, ${^6}$Li(p,$\gamma$)${^7}$Be, ${^6}$Li(p,$\alpha$)${^3}$He, ${^7}$Li(p,$\alpha$)${^4}$He, ${^7}$Li($\alpha,\gamma$)$^{11}$B, ${^7}$Be(n,p)${^7}$Li, ${^7}$Be(p,$\gamma){^8}$B, ${^7}$Be($\alpha,\gamma){^{11}}$C, ${^9}$Be(p,$\gamma){^{10}}$B, ${^9}$Be(p,d$\alpha){^4}$He, ${^9}$Be(p,$\alpha){^6}$Li, ${^9}$Be($\alpha$,n)$^{12}$C, $^{10}$B(p,$\gamma){^{11}}$C, $^{10}$B(p,$\alpha){^7}$Be, $^{11}$B(p,$\gamma){^{12}}$C, $^{11}$B(p,$\alpha\alpha){^4}$He, $^{12}$C(p,$\gamma){^{13}}$N, $^{12}$C($\alpha,\gamma){^{16}}$O, $^{13}$C(p,$\gamma){^{14}}$N, $^{13}$C($\alpha$,n)$^{16}$O, $^{13}$N(p,$\gamma){^{14}}$O, $^{14}$N(p,$\gamma){^{15}}$O, $^{15}$N(p,$\gamma){^{16}}$O and $^{15}$N(p,$\alpha){^{12}}$C \cite{Mi12}. In the present work we have employed the most recent values of neutron lifetime and the baryon-to-photon ratio and further modified $^3$He($^4$He,$\gamma$)$^7$Be reaction rate which is used directly for estimating the formation of $^7$Li as a result of $\beta^+$ decay \cite{Du17}. We have also used the most recent parametrization for t($^4$He,$\gamma$)$^7$Li and d($^4$He,$\gamma$)$^6$Li reaction rates in the temperature ranges up to 5$T_9$ \cite{Du17}, though it is found that these two modifications produced little effect on $^7$Li abundance over the earlier ones. We have also compared results of the present calculations with our previous one \cite{Mi12} and other recent calculations \cite{Coc14,Coc15,Cy16}, the most recent one \cite{Cy16} used the same values of $\tau_n$ and $\eta$ as our present work and performed calculations based on the PARTHENOPE code.     
\noindent
\subsection{ Impact of fundamental constants on the primordial nucleosynthesis }
\label{subsection5a}

    Apart from the astrophysical nuclear reaction rates, the primordial abundances depend on three additional parameters, the number of light neutrino flavours, the neutron lifetime and the baryon-to-photon ratio in the universe. In standard form the number of light neutrino flavours $N_\nu$ is taken as 3.0. The observations by the WMAP \cite{WMAP,WMAP1} and the Planck \cite{Planck,Planck1} space missions enabled precise extraction of the baryon-to-photon ratio of the Universe as $\eta=6.0914\pm0.0438 \times 10^{-10}$. The most recent experimental value for the neutron lifetime $\tau_n$ which is $880.3 \pm 1.1$ s \cite{Ol14} has been used in the present calculations. 

\begin{table*}[htbp]
\vspace{0.0cm}
\centering
\caption{\label{tab:table1} Yields at CMB-WMAP baryonic density ($\eta_{10}=6.0914 \pm 0.0438$ ~\cite{WMAP,WMAP1}).}
\begin{tabular}{ccccccc}
\hline
\hline
 &Ref. \cite{Mi12} (2012)&Ref. \cite{Coc14} (2014)&Ref. \cite{Coc15} (2015)&Ref. \cite{Cy16} (2016)&This work&Observations \\ \hline
 
${^4}$He&0.2479&0.2482$\pm$0.0003&0.2484$\pm$0.0002&0.2470&0.2467$\pm$0.0003&0.2449$\pm$0.0040~\cite{Aver15} \\
D/H$~(\times 10^{-5})$&2.563&2.64$^{+0.08}_{-0.07}$&2.45$\pm$0.05&2.579& 2.623$\pm$0.031&2.53$\pm$0.04~\cite{Cooke14} \\
${^3}$He/H$~(\times 10^{-5})$&1.058&1.05$\pm$0.03&1.07$\pm$0.03&0.9996&1.067$\pm$0.005&1.1$\pm$0.2~\cite{Ba02} \\
${^7}$Li/H$~(\times 10^{-10})$&5.019&4.94$^{+0.40}_{-0.38}$&5.61$\pm$0.26&4.648&4.447$\pm$0.067&1.58$^{+0.35}_{-0.28}$~\cite{Sb10} \\ \hline
\hline
\end{tabular} 
\vspace{-0.2cm}
\end{table*}
\noindent     

\noindent
\subsection{ Thermonuclear reaction rate for radiative ${^3}$He${^4}$He capture }
\label{subsection5b}

    The twelve most important nuclear reactions which affect the predictions of the abundances of the light elements [${^4}$He, D, ${^3}$He, ${^7}$Li]  are n$-$decay, p(n,$\gamma$)d, d(p,$\gamma){^3}$He, d(d,n)${^3}$He, d(d,p)t, ${^3}$He(n,p)t, t(d,n)${^4}$He, ${^3}$He(d,p)${^4}$He, ${^3}$He($\alpha,\gamma){^7}$Be, t($\alpha,\gamma){^7}$Li, ${^7}$Be(n,p)${^7}$Li and ${^7}$Li(p,$\alpha){^4}$He. The uncertainties for the reactions ${^3}$He $+$ ${^4}$He $\rightarrow {^7}$Be $+$ $\gamma$, ${^3}$H $+$ ${^4}$He $\rightarrow {^7}$Li $+$ $\gamma$ and p $+$ ${^7}$Li $\rightarrow {^4}$He $+$ ${^4}$He directly reflect uncertainty in the predicted yield of ${^7}$Li.
    
    The ${^3}$He(${^4}$He,$\gamma$)${^7}$Be reaction rate which is used directly for estimating the formation of ${^7}$Li as a result of $\beta^+$ decay is now replaced by the most recent rate equation \cite{Du17}. The refined variant of calculations for the astrophysical S-factor of the d(${^4}$He,$\gamma$)${^6}$Li, t(${^4}$He,$\gamma$)${^7}$Li and ${^3}$He(${^4}$He,$\gamma$)${^7}$Be reactions, are in better agreement with the previously available as well as the most recent experimental data and its predictive reliability is also demonstrated. The new parametrization for the reaction rates \cite{Du17} given, respectively, by

\vspace{-0.25cm}    
\begin{eqnarray}
 N_A<\sigma v> =&&17.128/T_9^{2/3}\exp(-7.266/T_9^{1/3}) \nonumber\\
                 &&\times(1.0-4.686~T_9^{1/3}+15.877~T_9^{2/3} \nonumber\\
                 &&-21.523~T_9+18.703~T_9^{4/3}-4.554~T_9^{5/3}) \nonumber\\
                 &&+53.817/T_9^{3/2}\exp(-6.933/T_9),  
\label{seqn6}
\end{eqnarray}
\noindent

\vspace{-0.25cm}
\begin{eqnarray}
 N_A<\sigma v> =&&2304.319/T_9^{2/3}\exp(-6.165/T_9^{1/3}) \nonumber\\
                 &&\times(1.0-25.706~T_9^{1/3}+74.057~T_9^{2/3} \nonumber\\
                 &&+28.460~T_9-61.303~T_9^{4/3}+19.591~T_9^{5/3}) \nonumber\\
                 &&+29.322/T_9^{3/2}\exp(-1.641/T_9),  
\label{seqn7}
\end{eqnarray}
\noindent

\vspace{-0.25cm}
\begin{eqnarray}
 N_A<\sigma v> =&&36807.346/T_9^{2/3}\exp(-11.354/T_9^{1/3}) \nonumber\\
                 &&\times(1.0-15.748~T_9^{1/3}+56.148~T_9^{2/3} \nonumber\\
                 &&+27.650~T_9-66.643~T_9^{4/3}+21.709~T_9^{5/3}) \nonumber\\
                 &&+44350.648/T_9^{3/2}\exp(-16.383/T_9)  
\label{seqn8}
\end{eqnarray}
\noindent    
are used in the temperature ranges up to 5$T_9$ for the astrophysical evaluations of ${^6}$Li,  ${^7}$Li and ${^7}$Be productions. 
  
\noindent 
\section{ Results and Discussion }
\label{section6}
    
    A comprehensive study of the effect of fundamental constants and nuclear reaction rates on primordial nucleosynthesis is performed. All these calculations described so far for the standard BBN and with the modified reaction rates \cite{An99,An04,Du17} are performed with the most recent experimental value for the neutron lifetime $\tau_n = 880.3 \pm 1.1$ s and the value of $\eta = \eta_{10} \times 10^{-10} = 6.0914 \pm 0.0438 \times 10^{-10}$ for the baryon-to-photon ratio. In Table-I the results of present calculations have been compared with our previous one \cite{Mi12} and other recent calculations \cite{Coc14,Coc15,Cy16}. The theoretical uncertainties quoted in the table arise out of experimental uncertainties in the magnitudes of $\tau_n$ and $\eta_{10}$. However, the other dominant source of uncertainty arises from the reaction rates which would certainly increase the theoretical uncertainties quoted in this work. We find that using the most recent values of fundamental constants and new reaction rate result in marginal decrease in helium mass fraction causing slight improvement than obtained previously in standard BBN calculations. On the contrary, the relative abundances of deuteron and ${^3}$He increase marginally, yet remaining within the uncertainties of experimental observations. It is also observed that the yield of ${^7}$Li slightly improves ($\sim 12\%$) (which was off by a factor of $\sim 3$ in the standard BBN calculation) over the standard BBN abundance. It is, therefore, suggestive that even with considerable nuclear physics uncertainties, most of these nuclear reactions have minimal effect on the primordial lithium abundance problem of BBN.
    
\noindent
\section{ Summary and conclusion }
\label{section7}

    In summary, the predictions of the standard BBN theory depend on the astrophysical nuclear reaction rates and on three additional parameters, the number of flavours of light neutrinos, the neutron lifetime and the baryon-to-photon ratio in the universe. The effect of the modification of thirty-five reaction rates on light element abundance yields in BBN was investigated earlier by us. In the present work we have replaced the neutron lifetime and baryon-to-photon ratio by most recent values and further modified the reaction rates d($^4$He,$\gamma$)$^6$Li, t($^4$He,$\gamma$)$^7$Li and ${^3}$He(${^4}$He,$\gamma$)${^7}$Be (which is used directly for estimating the formation of ${^7}$Li as a result of $\beta^+$ decay) by the most recent equations \cite{Du17}. We have studied light element abundance yields as functions of evolution time or temperature. We find that these changes cause only slight improvement ($\sim 12\%$) on the standard BBN abundance yield of ${^7}$Li. In a few other recent studies \cite{Coc14,Coc15,Cy16,Bo10} also it was found that addition of some new reaction rates to the BBN code and thus increasing the reaction network had virtually no effect on the BBN abundances. It is interesting to note that the earlier observed relative abundance of $^7$Li ($1.1\pm0.1 \times 10^{-10}$) \cite{Ho09} has been revised upward by $\sim 44\%$ recently \cite{Sb10}. Moreover, if one takes the lower limit of the present theoretical estimate and compares it with the upper limit of the observed value of $^7$Li relative abundance then these two values appear to be converging but still overestimated by a factor of 2.27. If the other dominant source of uncertainty from the reaction rates is also considered then certainly the theoretical and the observed values would converge further. Nevertheless, the chances of solving either of the `lithium problems' by conventional nuclear physics means are unlikely and, if these problems remain up to future observations, we may be forced to consider more exotic scenarios.
    
    The motivations to extend BBN beyond the standard model could be to use it to probe the early universe and test fundamental physics as well as to find a solution to the lithium problem. If gravity differs from its general relativistic description, the rate of expansion of the universe may be affected and the variation of the fundamental constants may have to be constrained by BBN \cite{Coc09,Coc12}. The $^7$Li abundance may be lowered by decay of a massive particle during or after BBN. Similar effect could also be obtained with negatively charged relic particles, like the supersymmetric partner of the $\tau$ lepton, that could form bound states with nuclei, lowering the Coulomb barrier and thus leading to the enhancement of nuclear reactions \cite{Ku13}. Other non–standard solutions to the lithium problem comprise of photon cooling \cite{Er12}, possibly combining particle decay and magnetic fields \cite{Ya14}.

\end{document}